\documentclass{optica-article}

\journal{opticajournal} % for journals or Optica Open

\articletype{Research Article}

\usepackage{lineno}
\begin{document}

\title{Single- and multi-layer micro-scale diffractive lens fabrication for fiber imaging probes with versatile depth-of-field}

\author{Fei He,\authormark{1} Rafael Fuentes-Dom\'inguez,\authormark{1} Richard Cousins,\authormark{2} Christopher J. Mellor,\authormark{3} Jennifer K. Barton,\authormark{4,5} and George S. D. Gordon\authormark{1,*}}

\address{\authormark{1}Optics and Photonics Group, University of Nottingham, Nottingham, NG7 2RD, United Kingdom\\
\authormark{2}Nanoscale and Microscale Research Centre, University of Nottingham, Nottingham, NG7 2RD, United Kingdom\\
\authormark{3}School of Physics and Astronomy, University of Nottingham, Nottingham, NG7 2RD, United Kingdom\\
\authormark{4}Wyant College of Optical Science, University of Arizona, Tucson, 85721, United States\\
\authormark{5}Biomedical Engineering, University of Arizona, Tucson, 85721, United States}

\email{\authormark{*}ezzgsg@exmail.nottingham.ac.uk} %% email address is required; see note below about the corresponding author designation

% use {asbstract*} to suppress the copyright line. Copyright information will be added in production

\begin{abstract*} 
Hair-thin optical fiber endoscopes have opened up new paradigms for advanced imaging applications \emph{in vivo}. In certain applications, such as optical coherence tomography (OCT), light-shaping structures may be required on fiber facets to generate needle-like Bessel beams with large depth-of-field, while in others shorter depths of field with high lateral resolutions are preferable. In this paper, we demonstrate a novel method to fabricate light-shaping structures on optical fibres, achieved via bonding encapsulated planar diffractive lenses onto fiber facets. Diffractive metallic structures have the advantages of being simple to design, fabricate and transfer, and our encapsulation approach is scalable to multi-layer stacks. As a demonstration, we design and transfer a Fresnel zone plate and a diffractive axicon onto fiber facets, and show that the latter device generates a needle-like Bessel beam with 350 \textmu m focal depth. We also evaluate the imaging performance of both devices and show that the axicon fiber is able to maintain focussed images of a USAF resolution target over a 150 \textmu m distance. Finally, we fabricate a two-layer stack of Fresnel zone plates on a fiber and characterise the modified beam profile and demonstrate good imaging performance. We anticipate our fabrication approach could enable multi-functional complex optical structures (e.g. using plasmonics, polarization control) to be integrated onto fibers for ultra-thin advanced imaging and sensing. 

\end{abstract*}

%%%%%%%%%%%%%%%%%%%%%%%%%%  body  %%%%%%%%%%%%%%%%%%%%%%%%%%
\section{Introduction}
\begin{figure}[ht!]
\centering\includegraphics[width=0.7\textwidth]{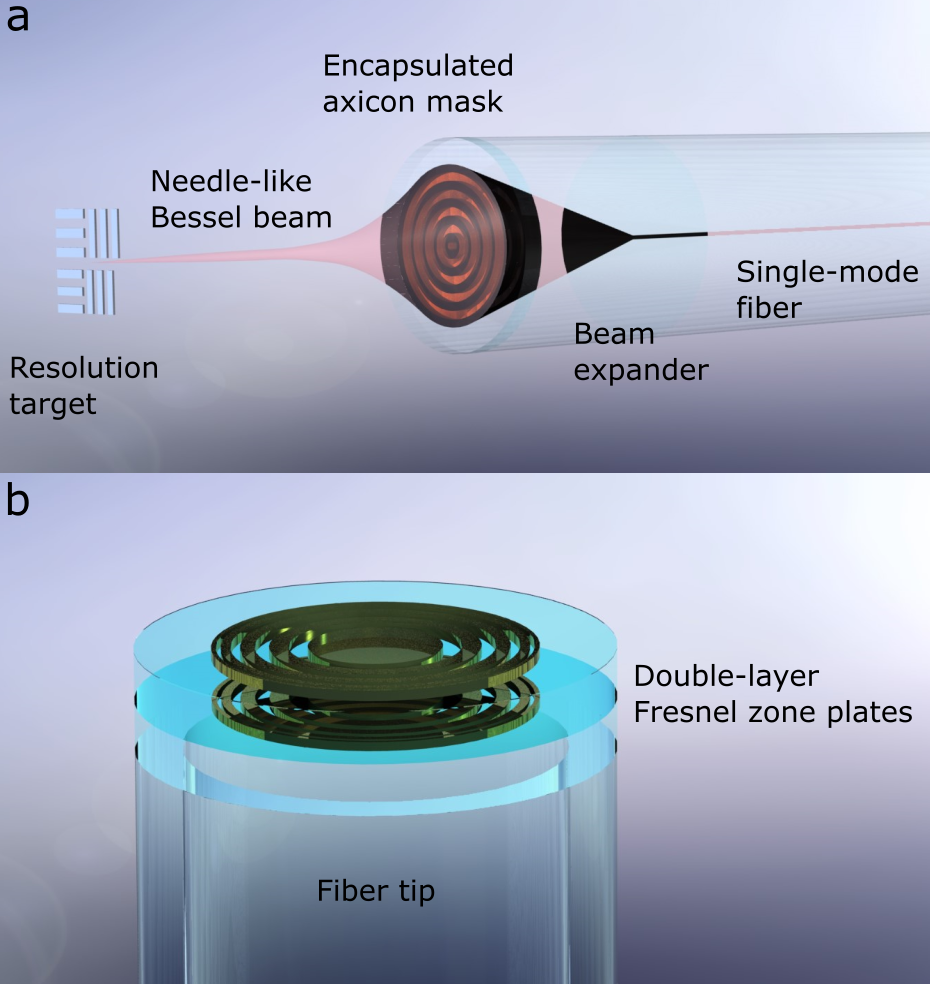}
\caption{Fiber endoscope devices fabricated by bonding encapsulated lenses onto fiber tips. (a) An encapsulated axicon mask bonded on top of a multi-mode fiber core (used for fiber-beam expansion) fusion spliced with a single-mode fiber. The fiber device is capable of focusing light into a needle-like Bessel beam and imaging a resolution target over a large depth range. (b) The fabrication process is also capable of transferring multi-layer lens stacks, such as double-layer Fresnel zone plates shown here, onto fiber tips.}
\label{fig:Fig1}
\end{figure}

Hair-thin fiber endoscopes are creating new paradigms for biological imaging applications with minimal invasiveness and sub-cellular resolution, such as high-resolution confocal microscopy \cite{loterie2015digital}; \emph{in vivo} bright/dark-field and fluorescence microscopy in living brains \cite{vcivzmar2012exploiting,vasquez2018subcellular}; quantitative phase/polarisation imaging for early-stage cancer detection \cite{gataric2018reconstruction,gordon2019characterizing}; and optical coherence tomography (OCT) in deep biological tissues \cite{pahlevaninezhad2018nano,wang2019miniature,li2020ultrathin}.

To properly perform functionalities desired by various imaging applications, fiber endoscopes may require on-tip light-shaping structures to manipulate the light flow. For instance, OCT applications require light focus with large depth-of-field (DOF) to trade off lateral and axial resolutions \cite{zhao2022flexible}, which can be enabled by shaping light into needle-like Bessel beams. They have desirable properties such as being non-diffracting (maintenance of the transversal intensity profile along large propagation length) and self-healing (recovery of original beam profile after the obstruction by finite-sized particles) \cite{mazilu2010light,khonina2021modern}. Therefore they are widely used in various applications such as nano-particle trapping \cite{liang2019direct,park2021optical,porfirev2021realisation}, laser machining \cite{nguyen2020non} and material processing \cite{bhuyan2010high,duocastella2012bessel}. In particular, their nearly intact lateral spot size maintained within the large DOF enables high-resolution tomography in deep biological tissues \cite{pahlevaninezhad2022metasurface,cao2022optical}.

Needle-like Bessel beams can be generated by refracting light through conical axicon lenses to uniformly spread the light along a large DOF \cite{mazilu2010light,khonina2021modern}, as opposed to tight Gaussian foci concentrated by spherical lenses \cite{gorelick2020axilenses}. This concept has been applied to make refractive axicon lenses on fiber endoscopes, which can be realised via focused-ion-beam (FIB) milling \cite{sloyan2020focused}, chemical etching \cite{kuchmizhak2014high,bachus2016fabrication,vairagi2017deep} and mechanical polishing \cite{wang2019miniature}. Similar light-shaping behavior can also be achieved via 3D printing photonic structures onto fiber tips \cite{gissibl2016two,reddy20223d,lightman2022vortex}. For more compact designs, 2D planar optics (i.e. diffractive lenses \cite{huang2018planar}) with sub-wavelength thickness can be designed with nano- and micro-scale structures that enable spatial control of amplitude, phase and polarization with subwavelength resolution\cite{chen2016review, kamali2018review}.

Designs for planar diffractive axicons are well established, for example binary-amplitude cosinusoidal gratings designed using computer-generated holography \cite{vasara1989realization,kolodziejczyk1990light,sochacki1992nonparaxial}. There are many possible ways to implement such planar diffractive lenses onto fiber tips \cite{yu2015optical,zhao2022optical}. These methods include, for example, either FIB milling the designed patterns into ultra-thin metallic layers deposited onto fiber tips \cite{principe2017optical,xomalis2018fibre,zeisberger2021plasmonic}, or directly printing nanostructures onto the fiber facets \cite{hadibrata2021inverse, plidschun2021ultrahigh, ren2022achromatic}. Separately, a number of techniques have been developed to transfer pre-fabricated patterns onto fibers \cite{smythe2009technique,lipomi2011patterning, sun2022quasi}. While these previous fabrication approaches all demonstrate effective light control at fiber facets, they are not all compatible with multi-layer stacking of devices or metasurfaces. This limits their applicability in emerging imaging applications that use multilayered structures to enable advanced imaging modalities \cite{avayu2017composite,gordon2019characterizing,zhang2023universal,kim2023multilayer,zhou2018multilayer}. Further, the imaging performance of diffractive lenses on fiber tips has rarely been evaluated. It is therefore critically important to evaluate whether aberrations in the point spread function allow acceptable quality imaging with these devices.

In this paper, we present a novel method to fabricate single- and multi-layer diffractive light-shaping structures on optical fibres. It is achieved via transfer bonding pre-fabricated diffractive lenses onto fiber facets, which means the devices are simple to design, fabricate and transfer. The fabrication process uses a conventional top-down method to define nano- and micro-structures on bulk substrates via electron-beam lithography (EBL) and lift-off processes, but is adapted to facilitate repeated transfer to optical fibers for multi-layer fabrication. Specifically, diffractive lenses are encapsulated in optically transparent polymers, detached from the substrate by etching a sacrificial layer, then finally bonded to fibre tips. As proof-of-principle, we demonstrate Fresnel zone plates and diffractive axicons bonded onto optical fiber facets shaping light into Gaussian foci and Bessel-like beams (Fig. \ref{fig:Fig1}a), and show that the axicon fiber is able to maintain focussed images of a USAF resolution target over a 150 \textmu m distance. We also show that our fabrication method is readily compatible with multi-layer lens stacks by fabricating and characterising a double-layer Fresnel zone plate structure on a fibre tip (Fig. \ref{fig:Fig1}b). The fabricated double-layer fiber device produces high quality images of a resolution target, demonstrating potential for future multifunctional devices (e.g. using polarization control, plasmonics) to be implemented on fibre tips.

\section{Methods}

\subsection{Fabrication Process}

\begin{figure}[ht!]    \centering\includegraphics[width=1.0\textwidth]{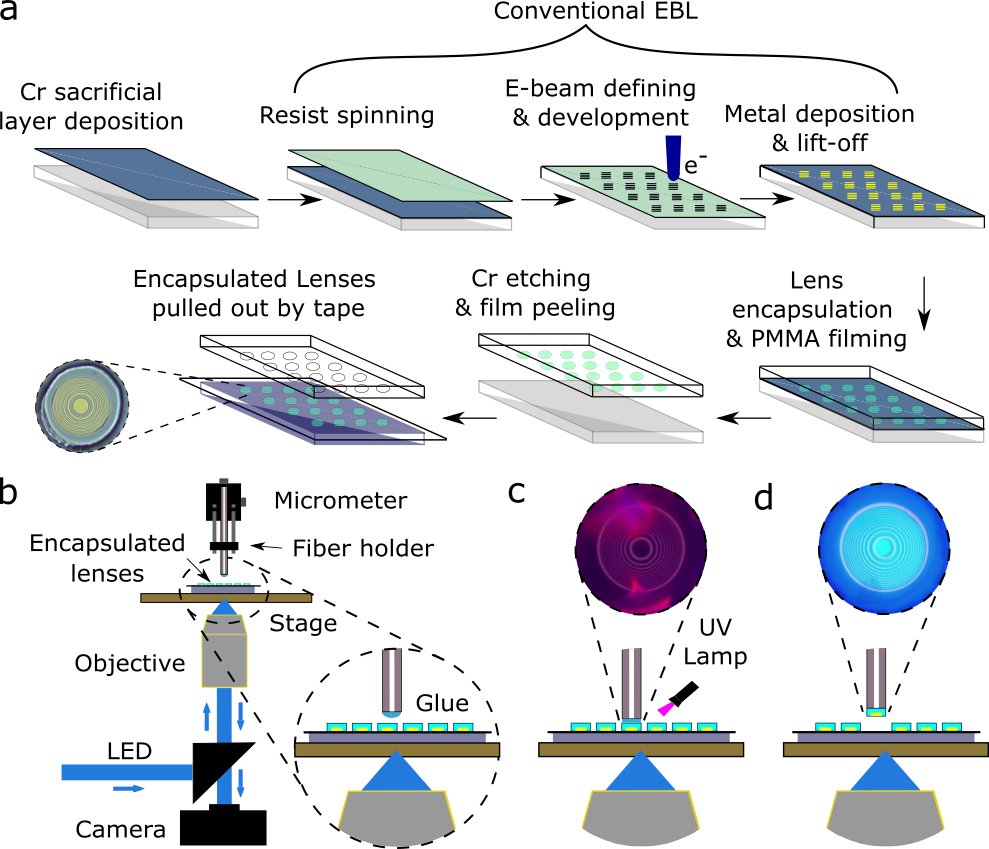}
\caption{Fabrication of ultra-thin polymer-encapsulated lenses onto fiber tips showing the steps: (a) to separate encapsulated lenses from the hosting substrate through etching of a Cr sacrificial layer and tape peeling, and (b-d) to align and adhere them onto optical fiber tips using an inverted microscope. (c) A glue-dipped fiber attached to a polymer-embedded lens on adhesive tape. An additional light source (i.e. white light) can be used to illuminate the fiber core to aid the alignment between the lens and the fiber. A UV lamp is used to cure the optical adhesive. (d) Post-curing, the encapsulated lens is adhered to the fiber tip and can be removed from the tape. The overlap between the light from the core and the lens illustrates that the cured adhesive is strong enough to maintain an excellent fiber-lens alignment during the removal process from tape.}
\label{fig:Fig2}
\end{figure}

The first fabrication steps follow a standard EBL procedure for metallic structures on silicon substrates using a single-stage lift-off process. As shown in Fig. \ref{fig:Fig2}a, a thin Cr sacrificial layer (5-10 nm) is deposited onto a silicon substrate, which is followed by spin-coating a positive tone e-beam resist (CSAR 6200.09), EBL and development processes to create lens patterns in the resist layer. Thermal evaporation is then used to deposit 100 nm Au (sufficient thickness for optical opacity), and the final lens patterns are defined by lift-off in AR 600-71. This highly standard first-step is a major advantage of our fabrication technique that allows easy fabrication of high-quality structures.

The next step is to spin-coat a 10-20 \textmu m layer of OrmoClear resist on top, align and pattern 125 \textmu m disks (matching the fiber diameter) via maskless photolithography thereby encapsulating the lenses in polymer. Next, a much thicker PMMA layer (500 \textmu m) is uniformly coated onto the Si substrate to assimilate the polymer-encapsulated lenses. Then the Cr sacrificial layer is dissolved via a wet etch process in a commercial chromium etchant, which releases the PMMA film with the gold pattern lenses. The polymer disks are then peeled away from the PMMA film using weakly adhesive tape on the lens-facing side. The tape containing the polymer disks is then fixed (via taping the edges) onto a glass coverslip which can be directly positioned onto the stage of a custom-built inverted optical microscope (as shown in Fig. \ref{fig:Fig2}b).

The inverted microscope allows imaging through the coverslip from underneath the stage in order to locate both lenses and fiber tips within the field of view. The live image at the tape-lens plane is formed by a camera capturing the reflection from illumination, allowing the selection of good-quality lenses and visualisation of fiber-lens alignment. An electromechanical stage enables 2D horizontal translation of the tape and lens samples, while a manual micrometer stage allows 3D movement of the fiber holder to align the encapsulated lenses with the fiber tip in the z axis. During this process, white light is sent through the fiber to help visualise the core area.

Before adhesion, the fibres must be prepared with a small beam-expanding section to allow the beam to fill the aperture of the diffractive lens. This is fabricated by fusion splicing a piece of MMF (FG105LCA) onto a single-mode fiber (SM600), and then cleaving the endcap to the desired length of 750 \textmu m. To adhere the polymer disks to the fiber, the fiber tip is firstly dipped in optical adhesive (NOA 68). Then, it is aligned with an encapsulated lens and lowered until the fiber is in contact with the polymer surface, indicated by a small amount of excess glue spreading out from the tip. As illustrated by Fig. \ref{fig:Fig2}c, the adhesive is cured via UV exposure for 30 -- 60 seconds. Due to the strong bond formed between the polymer disc and the tip, the encapsulated lens can be easily removed from the weakly adhesive tape by pulling the fiber up with the micrometer. The inset image in Fig. \ref{fig:Fig2}d demonstrates an excellent fiber-lens alignment maintained during the pulling process, as the light from the fiber core (the white-bright area) overlaps well with the lens pattern. We find the transfer and adhesion processes to be highly repeatable, typically only requiring 1-2 attempts to successfully make single- or multi-layer designs.  Re-fabrication after failed attempts is also possible, which further increases yield.

\subsection{Device Design and Simulation}
To demonstrate the versatility of our approach we designed two types of diffractive lenses with different focal properties: a Fresnel zone plate and a diffractive axicon. The Fresnel zone plate is designed with a focal distance of 250 \textmu m for 660 nm light, including accounting for the divergent incident wavefront caused by the fiber beam expander (a 750 \textmu m long and 105 \textmu m core-diameter multi-mode fiber segment fusion spliced onto the single-mode fiber \cite{plidschun2021ultrahigh}) that enables filling of the lens aperture. The design principle considers the optical path lengths from not only the rings to the focal point but also the single-mode fiber core to the rings \cite{kim2009fabrication}. For simplification, the multi-mode fiber (MMF) core is modeled as a homogeneous medium with uniform refractive index and the presence of the polymer encapsulation layer is neglected, assuming a zero-thickness binary mask on the homogeneous substrate surface. The diffractive lens can then be described by the following equation:
\begin{equation}
n \sqrt{L^2+R_m^2} + \sqrt{f^2+R_m^2} = nL+f+m \lambda/2 \quad (m = 1,2,...,N)
\end{equation}
where $f$ is the focal length of the fiber device, $L$ is the length of the beam expander (Fig. S1a of the supplementary), $n$ is the refractive index of the MMF core (1.46), \(\lambda\) is the wavelength of light, \(R_m\) denotes the inner/outer edges of the annular rings at an integer \(m^{th}\) order, and $N$ is the total number of ring edges which is 23 in this work.

The axicon pattern is created via a binarization process similar to previously published works \cite{vasara1989realization,kolodziejczyk1990light,sochacki1992nonparaxial}, which adjusts ring width and spacing to diffract input light into Bessel beam. Specifically in this work, the designed pattern retrieves the desired phase of a conical-shape wavefront from a spherical wave input (due again to the beam expander). In order to determine the correct profile of the amplitude grating, it is essential to know the phase profiles of the illumination beam and the output Bessel beam. The calculation of the illumination beam phase \(\psi_r\) at the MMF facet is illustrated in Supplementary Fig. S1b. The output phase of the conical wavefront \(\phi_r\) desired for the Bessel beam is represented by the slope angle of the wavefront across the radial coordinates \cite{chen2017generation} 
\begin{equation}
\phi_r= -\frac{2\pi}{\lambda}rsin\beta
\end{equation}
The slope angle is chosen as \(3.5^{\circ}\) to set the length and position of the needle to a reasonable range for optical depth imaging (<1mm, e.g. OCT) balanced with the practical fabrication requirement not to have too many or too few rings. Using geometrical optics, the needle length may be estimated using the formula $l = \frac{D}{2tan\beta}$ with $\beta = 3.5^{\circ}$ and lens diameter $D = 108$ \textmu m giving a value of $l \approx 890$ \textmu m, after which the light field begins to diverge significantly. Next, phase profile and the spherical wavefront correction term are combined to give a complex retardation:
\begin{equation}
T_r = \exp[i(\phi_r-\psi_r)]
\end{equation}
This is then normalised through its real component:
\begin{equation}
T_{norm} = \frac{T_r-min[\mathcal{R}(T_r)]}{max[\mathcal{R}(T_r)]-min[\mathcal{R}(T_r)]}
\end{equation}
Finally, the complex-valued mask function is binarized by the formula
\begin{equation}
T(x,y)= 
\begin{cases}
    1,&|T_{norm}|\geq0.85\\
    0,&|T_{norm}|<0.85
\end{cases}
\end{equation}
The generated pattern is illustrated in the schematic picture of Supplementary Fig. S1b. Compared to the Fresnel zone plate, the axicon mask consists of more evenly spaced rings, closer to a uniform annular diffraction grating. These binary diffractive lenses are simulated by Lumerical FDTD using full-wave analysis, assuming 100 nm thick gold masks (the same material and thickness used in the fabrication process) on homogeneous substrates (with the same refractive index as the MMF endcap core). Also, the presence of the encapsulation layers (1.54 refractive index and 15 \textmu m thickness, close to the fabricated thickness) are considered, giving close estimation to the output beam characteristics.

\subsection{Experimental Characterisation Setups}

\begin{figure}[ht!]
\centering\includegraphics[width=1.0\textwidth]{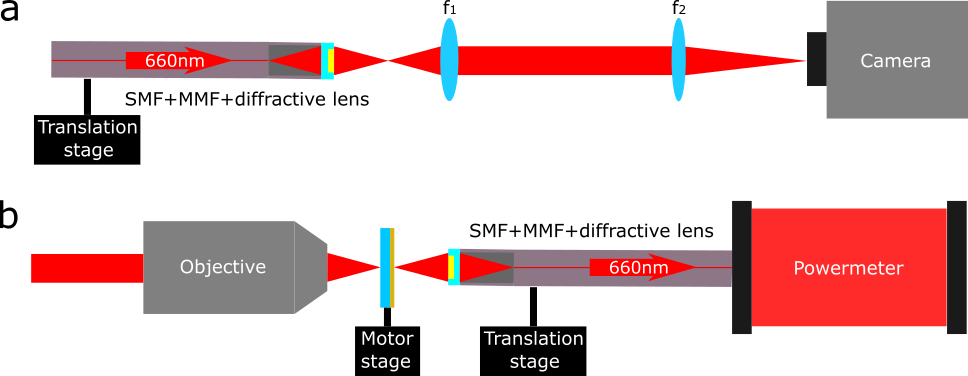}
\caption{Experimental optical setups for (a) output beam characterisation and (b) imaging performance test of the fiber devices. (a) The fiber patch cables are connected with a 660 nm laser source that illuminates the diffractive masks. A pair of lenses is used to form magnified focal spots onto a camera, and the depth-scan of the focal spot is accomplished via z-scan with a translation stage. (b) An objective (i.e. 10x and/or 20x) is used to focus the light onto a standardized resolution target, which is collected via the diffractive lenses into the fiber connected with a powermeter to record light intensity. A motor stage is used to laterally translate the target forming 2D images, and the depth scanning is performed similarly to (a) using z-translation.}
\label{fig:Fig3}
\end{figure}

To measure the output beam characteristics, the fiber devices are connected to a diode laser source (ThorLabs LPS-660-FC, 660 nm central wavelength), as illustrated in Fig. \ref{fig:Fig3}a. A pair of lenses (\(f_1\) = 18 mm and \(f_2\) = 250 mm, giving a magnification of 13.9), are used to form magnified images of the focal spots onto a camera (XIMEA MQ042RG-CM). The depth-scan is performed by z-translation of the fiber devices with a motorised stage. The imaging performance is tested by illuminating resolution targets in a transmission-mode configuration and collecting light with the fiber devices (Fig. \ref{fig:Fig3}b). An objective is used to focus a collimated light beam (from the 660 nm laser source) onto a standardized USAF resolution target (Thorlabs R1DS1N), which is collected through the fiber devices, which are connected in turn to an optical power meter (Thorlabs PM100USB). The resolution target is held and laterally translated via a motor stage to form 2D images, and the depth-scan is accomplished via a similar z-translation process. 

Uniform and focused illuminations are both widely used for fibre imaging experiments. In this work, we choose the latter method, based on related work in literature \cite{ren2022achromatic}, to emulate a real biomedical imaging scenario in which light is simultaneously focused and collected by the same fiber device in a reflection configuration. Further, because the resolution target is a planar transmissive pattern, rather than for example a highly scattering tissue sample, collimated incident light is not highly divergent after passing through the sample.  In this scenario, the fiber devices do not strongly exhibit their different depth-imaging abilities and behave more like `pin-hole' cameras with their small apertures not collecting out-of-focus light, regardless of the design of the lenses on their tips. Focussed illumination is therefore required to properly examine the difference in imaging performance between the different designs.

The imaging result may be different depending on the choice of illumination objectives numerical aperature (NA). A 20x objective (0.5 NA, 0.55 \textmu m FWHM spot size in glass) is used in this experiment because it is adequate to differentiate the depth-imaging abilities of the zone plate and axicon fibers. Low-NA objectives (i.e. 5x and 10x) may give better coupling efficiency of light due to their close match of NA to the fiber devices (\(<\) 0.2 NA) but reduce the ability to examine depth-imaging. We also note that the imaging experiment is achieved via a transmission measurement, the result of which can be different to the measurements in reflection. As light illumination and collection are simultaneously achieved by the fiber device in a reflection measurement, its point-spread-function has a double influence on the retrieved images.

In order to quantify imaging performance, the recorded images are compared with an ideal in focus `ground truth' image of the resolution target.  Specifically, a 2D correlation measurement is used:
\begin{equation}
corr = \frac{\sum_{x}\sum_{y}(Ig_{xy}-\overline{Ig})(Ir_{xy}-\overline{Ir})}{\sqrt{(\sum_{x}\sum_{y}(Ig_{xy}-\overline{Ig})^2)(\sum_{x}\sum_{y}(Ir_{xy}-\overline{Ir})^2)}}
\end{equation}
where \(Ig_{xy}\) and \(Ir_{xy}\) are the values of each pixels in the ground-truth and fiber-retrieved images respectively, and \(\overline{Ig}\) and \(\overline{Ir}\) are the mean values of the images.

We note that fringe contrast and edge response calculations are also widely used to analyse imaging performance of resolution targets. However, beyond a certain distance from the fibre/lens (e.g. the bare-fiber images after 200 \textmu m distance) it is difficult to distinguish peaks and valleys so the fringe contrast cannot be accurately estimated. Similarly, artefacts present in the background, due to additional speckle caused by the diffractive mask and sample imperfections, make edge response computations more susceptible to noise. The similarity approach is therefore more robust and gives the most reliable performance comparison. 

\section{Results}

\subsection{Device Beam Simulation and Characterisation}

\begin{figure}[ht!]
    \centering    \includegraphics[width=1.0\textwidth]{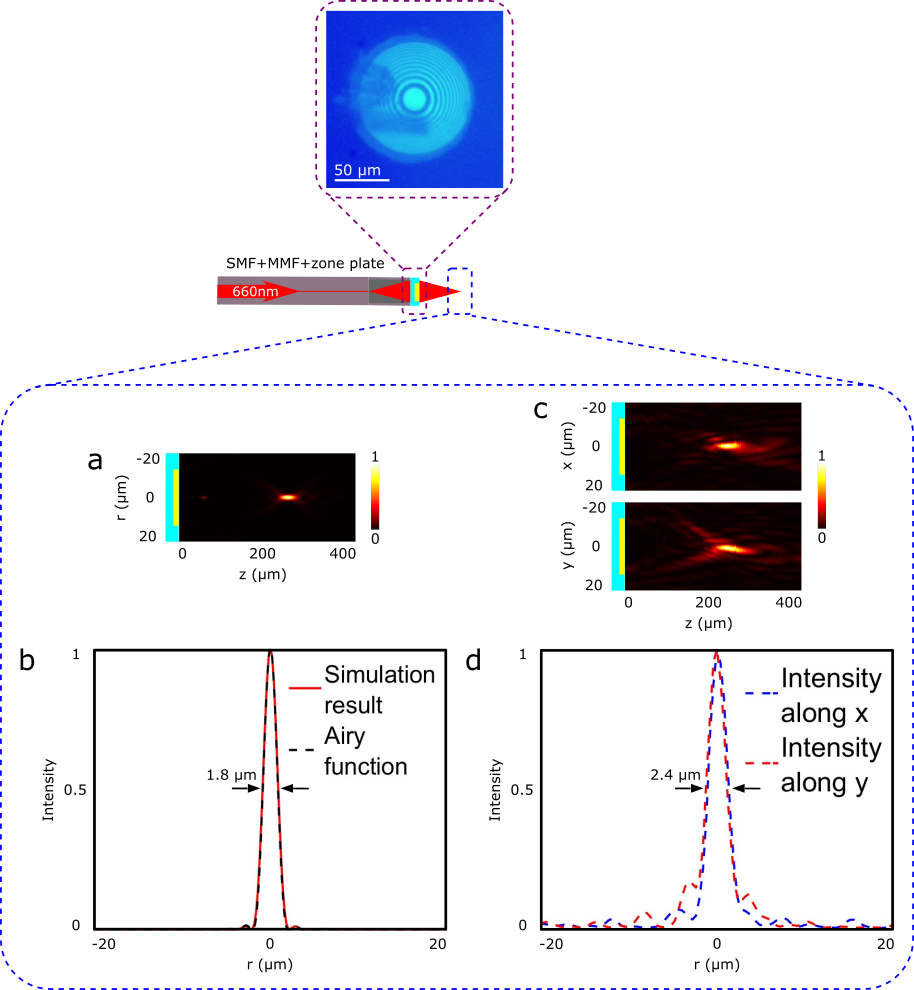}
    \caption{Comparison between the numerically simulated (a-b) and experimentally measured (c-d) output field characteristics from the Fresnel zone plate fiber device shaping light into Gaussian foci. (a) and (c) show the cross-sectional intensity maps of the Gaussian foci. (b) and (d) show the line profiles of the Gaussian foci at the peak field-intensity planes, with z=245 \textmu m in (a) and z=235 \textmu m in (c) respectively. The result in (b) is overlaid with theoretical Airy function. The microscope image of the fiber-attached zone plate is shown in the inset.}
    \label{fig:Fig4}
\end{figure}

We first simulate the performance of the output fields from the fiber devices, and then compare it with the experimental measurement result. The simulation result in Fig. \ref{fig:Fig4}a shows that the zone plate tightly focuses the light with intensity peak at 245 \textmu m and 70 \textmu m depth-of-field. The lateral intensity profile in Fig. \ref{fig:Fig4}b further illustrates that light is shaped into a high-quality Gaussian focus with a 1.8 \textmu m full width of half maximum (FWHM), which is also indicated by a close match between the simulation and the theoretical Airy function:
\begin{equation}
I = \left[\frac{2J_1(k \cdot NA \cdot r)}{k \cdot NA \cdot r}\right]^2
\end{equation}
where \(J_1\) is a Bessel function of the first kind, and the numerical aperture (NA) of the zone plate can be directly calculated using the focal distance and the lens diameter (100 \textmu m). The experimentally measured output light field from the zone plate fiber (via the setup in Fig. \ref{fig:Fig3}a) shows that the fabricated fiber device has output beam characteristics well-suited for fiber-imaging applications. This is illustrated in Fig. \ref{fig:Fig4}c by the tight focus at 235 \textmu m with 100 \textmu m depth-of-field, a close match to the simulation result. The lateral intensity profiles in Fig. \ref{fig:Fig4}d further demonstrate that light is shaped into a high-quality Gaussian focus with a small degree of aberration and a 2.4 \textmu m full width of half maximum (FWHM) spot size. The slight shift on the focal distance as well as the aberration on the focus can be attributed to mechanical distortion during the fabrication process and endcap length variation during the fiber-cleaving process. 

\begin{figure}[ht!]
    \centering    \includegraphics[width=1.0\textwidth]{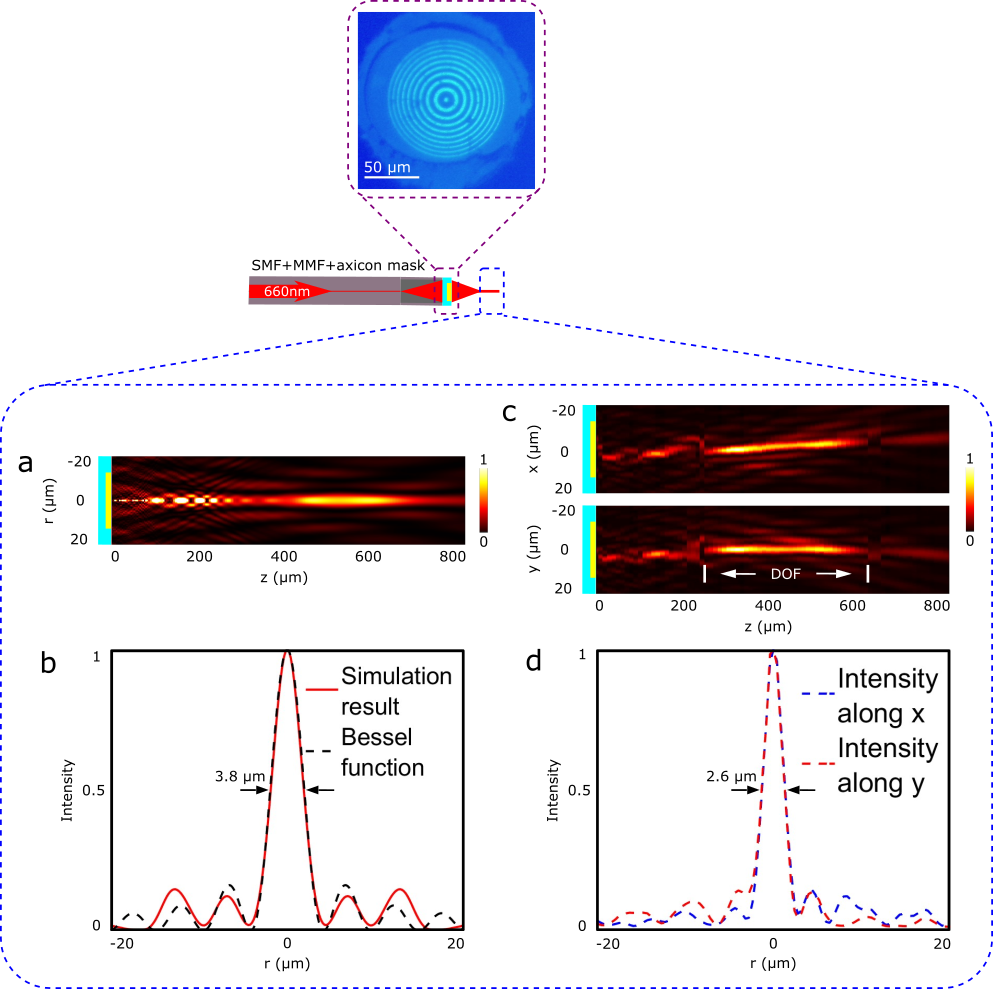}
    \caption{Comparison between the numerically simulated (a-b) and experimentally measured (c-d) output field characteristics from the axicon fiber device shaping light into Bessel beams. (a) and (c) show the cross-sectional intensity maps of the Bessel beams. (b) and (d) show the line profiles of the Bessel beams at the peak field-intensity planes, with z=450 \textmu m in (b) and z=320 \textmu m in (d) respectively. The result in (b) is overlaid with theoretical Bessel function. The microscope image of the fiber-attached axicon mask is shown in the inset.}
    \label{fig:Fig5}
\end{figure}

In contrast to the zone plate fiber, the axicon fiber shapes light into a needle-like Bessel beam through the overlap between diffraction orders from the annular grating (Fig. \ref{fig:Fig5}a). Due to the diffractive nature of the mask, the needle is not purely continuous and instead displays discrete high-order diffraction lobes at intermediate distances before the main lobe. The main lobe has a length of 400 \textmu m, and the light field diverges significantly after 800 \textmu m distance from the fiber facet, closely matching the predictions from geometrical optics. The lateral intensity profile in Fig. \ref{fig:Fig5}b further illustrates that the field has zero-order Bessel function distribution (as indicated by the pronounced side lobes) with a 3.8 \textmu m FWHM. The simulated result is also overlaid with the theoretical Bessel function:
\begin{equation}
I = [J_0(k \cdot NA \cdot r)]^2
\end{equation}
where \(J_0\) is the Bessel function of the zeroth order and the NA is directly related to the designed slope angle \(3.5^{\circ}\) mentioned in the device design section. The experimental measurement result in Fig. \ref{fig:Fig5}c illustrates that the output light maintains a needle-like shape for a distance of 650 \textmu m before the light begins to diverge significantly, and that the main lobe has 350 \textmu m depth-of-field (DOF). The lateral intensity profile in Fig. \ref{fig:Fig5}d further proves that the field has zero-order Bessel function distribution with a 2.6 \textmu m FWHM spot size. The measured spot size and the needle range differ somewhat from the simulated result and the measured properties suggest that they are produced by an axicon with an equivalent slope angle of \(5^{\circ}\), slightly larger than designed. This deviation can be attributed to imperfections including damage of the axicon mask and distortion of the polymer layer during the fabrication process. Despite these imperfections, a tight lateral spot is maintained over a large depth-of-focus, making the axicon-fiber device a good candidate for depth imaging, such as fiber-based OCT.

\subsection{Experimental Imaging Performance}

\begin{figure}[ht!]
    \centering    \includegraphics[width=1.0\textwidth]{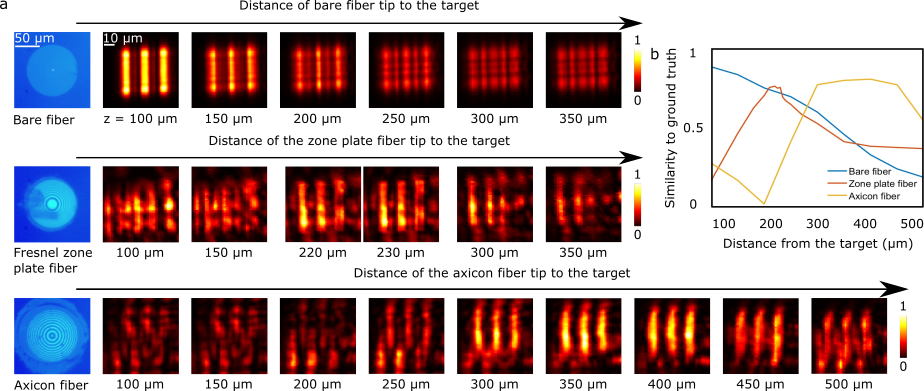}
    \caption{Imaging performance of the fiber devices. (a) Images of the resolution target retrieved from a bare fiber and the zone plate and axicon fibers at varying distances to the target. The images have an identical size of 60 \textmu m by 60 \textmu m in x \& y axes. (b) Similarities of the recorded images to the ground truth image at varying distances from 100 \textmu m to 500 \textmu m calculated via 2D-correlation.}
    \label{fig:Fig6}
\end{figure}

Next, we evaluate our two fiber devices in a resolution target imaging experiment. This shows the achievable imaging quality of the different designs and thus their expected performance in imaging applications. For endoscopic imaging applications, the ability of these devices to image at different working distances is important. The characterisation of different working distances is achieved via the setup shown in Fig. \ref{fig:Fig3}b using a 20x objective with 0.5 NA which is selected so as to emulate a diverging beam from the target as would be expected in typical endoscopic reflection-mode imaging. The scanned resolution pattern has a width of 7.8 \textmu m (64 pairs/mm, group 6 element 1). As illustrated in Fig. \ref{fig:Fig6}a, a bare fiber tip (SM600 single-mode fiber) is able to retrieve a high-fidelity image but only at small distances ($<$ 100 \textmu m from the target). However, the images become significantly blurred at $>$ 150 \textmu m, and the main features of the pattern are unrecognisable after 200 \textmu m. Overall, the quality of the retrieved images drops dramatically as the fiber moves away from the target. This observation is quantitatively analysed in Fig. \ref{fig:Fig6}b by the blue line -- the correlation between the ground truth image and the fiber-retrieved images. 

Despite the presence of some noise, the Fresnel zone plate fiber device is able to retrieve high-quality images (0.76 similarity) at a distance of around 230 \textmu m, the device focal point, as indicated by the orange line in Fig. \ref{fig:Fig6}b. Compared with bare fiber, the zone plate fiber has the advantage of retrieving a high-fidelity image at a distance that a bare fiber cannot, although with some additional artefacts that appear as background speckle. These artefacts may be due to the slightly aberrated point-spread-function caused by fabrication imperfections on the lens, such as the shadow area in the inset picture of Fig. \ref{fig:Fig4}. There may also be further artefacts arising from edge diffraction on the resolution target, though these could be reduced in future by using a less coherent light source such as an LED.

We also note that zone plate fiber devices can have higher imaging resolutions than a bare fiber. For instance, the FWHM spot size would give a 2.4 \textmu m resolution for this device according to the Rayleigh criterion, which is much smaller than the core size of the single-mode fiber. Such devices can be designed to retrieve images at a large range of distances, at the expense of resolution requirements. Here, the possible range of focal lengths is from a few microns (limited by ensuring ring sizes are larger than the EBL limitation, i.e. 20 nm) to a few millimeters (limited by having enough rings on a fibre facet to achieve focussing). 

The axicon fiber is able to retrieve high-quality images between 300 \textmu m to 450 \textmu m distances, as shown in Figure \ref{fig:Fig6}a.  Fig. \ref{fig:Fig6}b shows that the similarity to the ground truth plateaus at 0.77 -- 0.8 over a 150 \textmu m depth, 7.5 times longer than that of the zone plate fiber. 

\subsection{Extension to Multiple Layers}

\begin{figure}[ht!]
    \centering    \includegraphics[width=1.0\textwidth]{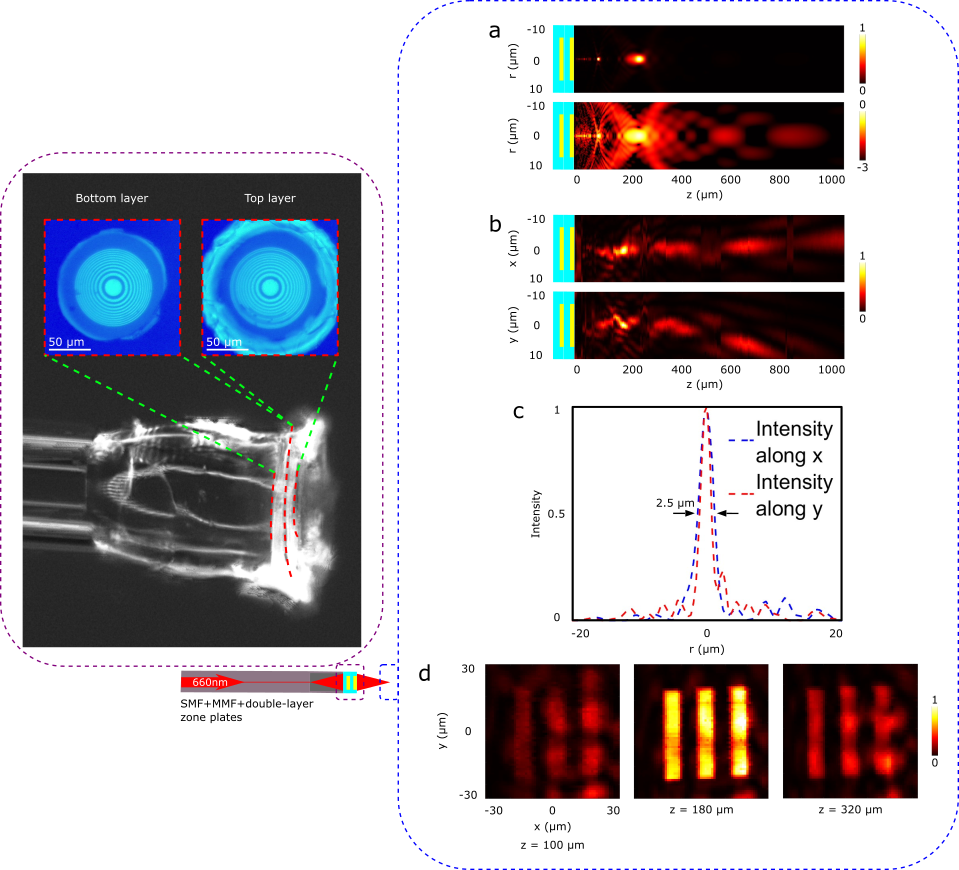}
    \caption{Extension of our fabrication method to double-layer lenses on fiber tip. (a) Numerical simulation of the output fields from a double-layer zone plate fiber device, showing the central-symmetrical foci in linear (top) and logarithm (bottom) scales. (b) Experimental characterisation of the output fields (linear scale) from fabricated double-layer fiber device, showing the cross-sectional intensity maps in the x-z and y-z planes. (c) Line profiles of the focal spot at the field intensity-peak plane, z=180 \textmu m in (b). (d) Images of the resolution target retrieved from the fiber device at the distances of z=100 \textmu m, 180 \textmu m and 320 \textmu m (3 focusing lobes). The microscope images of the fiber device are shown in the inset.}
    \label{fig:Fig7}
\end{figure}

Finally, we show the compatibility of our fabrication method with multi-layer structures by fabricating a double-layer Fresnel zone plate on a fiber. The device has two layers each of which contains an identical zone plate (same design as the single-layer presented in section 2.2), and is fabricated by performing the disk transfer and adhesion process twice in sequence (Fig. \ref{fig:Fig2}b--d). As shown in the inset of Fig. \ref{fig:Fig7}, minor imperfections can be identified on the fabricated device especially the top layer -- one missing ring together with some damage of other rings during the detachment from the hosting tape. Such imperfections may affect quality of the output field, leading to differences between the experimental characterisation and numerical simulation. 

The simulation result shows that the double-layer device focuses light at 240 \textmu m (slightly aberrated) with an equally strong side lobe at closer distance to the fiber tip (Fig. \ref{fig:Fig7}a, top panel). The focal point does not have significant shift compared with that of the single-layer zone plate device, because the two diffractive masks have limited separation (15 \textmu m, the polymer thickness). This means that the two zone plates can be considered as one connected mask, giving insignificant positional shift on the main focusing lobe. However, due to the tiny perturbation on the input field by the second mask, an equally strong side lobe is created and the intensity distribution in the main lobe is slightly asymmetrical. The perturbation may be further enlarged due to any imperfection appearing on the masks, such as our fabricated double-layer zone plates. 

The experimentally characterised output field shows that the main focus lobe is shifted to 180 \textmu m (60 \textmu m closer to the facet compared with the simulation result), and has noticeable additional lobes at larger distances (Fig. \ref{fig:Fig7}b). The existence of these additional lobes is validated by the simulated field map in logarithm scale (Fig. \ref{fig:Fig7}a, bottom panel), but they are not properly suppressed due to the fabrication imperfection of the double-layer device. Although the lateral focal spot has noticeable aberration (illustrated by the slightly asymmetrical spot shape and side diffraction rings in Fig. \ref{fig:Fig7}c), the central spot still has a clear Gaussian distribution, which indicates that the fiber device can have good imaging performance. 

This is illustrated in Fig. \ref{fig:Fig7}d, in which images are taken via the setup shown in Fig. \ref{fig:Fig3}b with the same resolution pattern and a 10x objective (0.3 NA, 0.9 \textmu m FWHM spot size in glass). These show the device is able to retrieve a high-quality image from the resolution target at the focal point z=180 \textmu m, and the pattern can also be be retrieved at two side focusing lobes (z=100 \textmu m and 320 \textmu m), albeit with compromised image contrast and quality. Due to the highly lossy nature of the double-layer diffractive lenses, using the high-NA objective (20x) further diminishes light coupling into the fiber. Therefore, in this experiment a 10x objective, which has a closer NA to the fiber device, is used to ensure adequate collection of light by the fiber device to enable high-fidelity images. An additional effect of this is that the images appear sharper due to the less divergent light.

\section{Discussion}

We have presented a method to fabricate diffractive lenses on fibers via polymer encapsulation and transfer using UV-curable adhesive. As proof-of-principle, we designed, fabricated and transferred a Fresnel zone plate and a diffractive axicon onto MMF-capped single-mode fibers. Compared to the zone plate fiber focusing light into a tight Gaussian focus, the axicon fiber device shapes light into needle-like Bessel beam with a 2.6 \textmu m (FWHM) lateral spot maintained over a 350 \textmu m depth-of-field. This shows the flexibility of this approach to create lenses with different focal profiles tailored to different applications. We also investigated their imaging capabilities, and showed that the fiber devices are able to retrieve high-quality images from a standarised resolution pattern although with some aberration. In particular, the axicon fiber is capable of retrieving high-quality images over a 150 \textmu m distance. Finally, we demonstrated that our fabrication method can readily be extended to fabricate multi-layer structures onto fibers, which could enable more complex multi-functional light-processing at the fibre tip. As proof-of-principle, we fabricated double-layer Fresnel zone plates onto fibers, and confirmed that the device has good imaging performance.

There are two main challenges we encountered in this work. The first challenge is the last step of the fabrication process -- removing the fiber-attached lenses from the hosting tape. Due to the sticky nature of tape, the pulling-up process creates mechanical force on the lens-polymer layer, which risks distorting the layer and the lens structure inside. In the worst case, this can impose noticeable damage to the lenses such as we observed in the top layer of the double-layer fiber device. However, fortunately the OrmoClear resist layer is strong enough to hold and protect the encapsulated structures in the majority of cases. This may be avoided in future by immersing the encapsulated lenses together with the hosting tape into solvents (i.e. acetone, IPA and toluene) before the fiber-sticking process to soften the sticky material on the tape, which would diminish the stretching force during the pulling-up process. The second challenge comes from the diffractive nature of the axicon mask, the output field of which necessarily contains discrete scattering lobes at short distances before the main lobe, as opposed to the desired continuous needle. This issue may be solved by designing discrete phase plate that overlaps multiple foci into a long and uniform needle \cite{cao2023optical}, or designing dielectric metalenses that precisely control the output amplitude and phase profile \cite{chen2017generation}. Dielectric metalenses also have the advantage of being transparent at optical frequencies, overcoming the efficiency shortage imposed by their metallic counterparts \cite{kamali2018review}.  However, our transfer and encapsulation processes may not be suitable for dielectric structures.

To put our work in context, it is important to compare with previously reported fiber-transfer methods from literature, as shown in Table. \ref{tab:table1}. Overall, we demonstrated a transfer process for metallic diffractive lenses which is compatible with multi-layer structures, and we also evaluated the imaging performance of three lens designs. 

Compared with the works of Lipomi and Kim \cite{lipomi2011patterning,kim2023multilayer}, our polymer layer has independent thickness to the encapsulated structure. This layer not only provides mechanical protection to the structure during transfer process, but also offers another degree of freedom to the structural design. For instance, these layers can be designed with different optical thicknesses providing separations between functional layers. The combination of them may give a functionality that is unachievable for a single-layer device, such as the works demonstrated by Avayu and Zhou \cite{avayu2017composite,zhou2018multilayer}. By stacking triple-layer plasmonic and/or dielectric metalenses onto fibers, multi-wavelength achromatic foci can be achieved, desirable for many multi-spectral imaging applications. Owing to the good imaging performance of our fiber devices, we anticipate that these devices could be used as widely-deployable endoscopic imaging devices such as fiber-OCT probes. They can be further expanded to designs such as spatio-temporal pulse-shaping axicons \cite{piccardo2023broadband} and super-oscillatory lenses generating sub-diffraction features \cite{yuan2014planar,qin2017supercritical,yuan2017achromatic}. Also, the micro-sized diffractive lenses may be fabricated via UV-lithography, making the manufacture process suitable to be scaled up. Finally but more importantly, the compatibility of our fiber devices to multi-layered designs could enable a new generation of multi-modality ultra-thin fiber-imaging devices, such as complex polarisation state converters or reflective beacons for fiber characterisation \cite{gordon2019characterizing} and metalens stacks for achromatic fiber imaging \cite{avayu2017composite,zhou2018multilayer}.

\begin{table}
    \centering
    \begin{tabular}{cccccc}
         \hline
         Author& Material & \multicolumn{1}{p{1.2cm}}{\centering Multi-layer} & \multicolumn{1}{p{1.2cm}}{\centering Fiber\\attached} & \multicolumn{1}{p{2cm}}{\centering Optical\\characterisation} & \multicolumn{1}{p{2cm}}{\centering Imaging\\inspection}\\
         \hline
         Smythe et al. \cite{smythe2009technique} & metallic & no & yes & no & no\\
         Lipomi et al. \cite{lipomi2011patterning} & metallic & yes & yes & no & no\\
         Kim et al. \cite{kim2023multilayer} & polymer & yes & yes & yes & no\\
         Sun et al. \cite{sun2022quasi} & metallic & no & yes & yes & not applicable\\
         Avayu et al. \cite{avayu2017composite} & metallic & yes & no & yes & yes\\
         Zhou et al. \cite{zhou2018multilayer} & dielectric & yes & no & yes & yes\\
         Zhang et al. \cite{zhang2023universal} & dielectric & to be done & to be done & yes & no\\
         \textbf{This work} & metallic & yes & yes & yes & yes\\
         \hline
    \end{tabular}
    \caption{Summary of previously reported methods to transfer nanostuctures onto fibers.}
    \label{tab:table1}
\end{table}

\begin{backmatter}
\bmsection{Funding}
This work is supported by UKRI Future Leaders Fellowship (MR/T041951/1).

\bmsection{Acknowledgments}
The authors would like to acknowledge the Nanoscale and Microscale Research Centre (nmRC) and School of Physics at the University of Nottingham for providing access to cleanroom facilities. 

\bmsection{Disclosures}
The authors declare no conflicts of interest.

\bmsection{Data availability} Data underlying the results presented in this paper are not publicly available at this time but may be obtained from the authors upon reasonable
request.

\bmsection{Supplemental document}
See Supplement 1 for supporting content. 

\end{backmatter}

%%%%%%%%%%%%%%%%%%%%%%% References %%%%%%%%%%%%%%%%%%%%%%%%%

%%%%%%%%%% If using BibTeX:

\end{document}